\documentclass[twocolumn]{aastex7}

\usepackage{amsmath}
\usepackage{orcidlink}
\usepackage{enumitem}

\defcitealias{beck2016}{B16}

\newcommand{\photoz}{photo-$z$}
\newcommand{\Photoz}{Photo-$z$}
\newcommand{\photozs}{photo-$z$'s}

\newcommand{\zphot}{$z_\mathrm{phot}$}
\newcommand{\specz}{spec-$z$}

\newcommand{\speczs}{spec-$z$'s}

\newcommand{\zspec}{$z_\mathrm{spec}$}
\newcommand{\nmad}{$\sigma_\mathrm{NMAD}$}
\newcommand{\sigmascatter}{$\sigma_\mathrm{scatter}$}
\newcommand{\sigmaatten}{$\sigma_\mathrm{atten}$}
\newcommand{\fout}{$f_\mathrm{outlier}$}

\newcommand{\meanbt}{$\langle$B/T$\rangle$}
\newcommand{\encap}{\texttt{encapzulate}}

\begin{document}

\title{Deep Learning Improves Photometric Redshifts in All Regions of Color Space}

\correspondingauthor{Brett H.~Andrews}
\email{andrewsb@pitt.edu}

\author[0009-0002-1352-8296]{Emma R.~Moran}
\affiliation{Department of Physics and Astronomy, University of Pittsburgh, Pittsburgh, PA 15260, USA}
\affiliation{Pittsburgh Particle Physics, Astrophysics, and Cosmology Center (PITT PACC), University of Pittsburgh, Pittsburgh, PA 15260, USA}
\email{erm160@pitt.edu}

\author[0000-0001-8085-5890]{Brett H.~Andrews}
\affiliation{Department of Physics and Astronomy, University of Pittsburgh, Pittsburgh, PA 15260, USA}
\affiliation{Pittsburgh Particle Physics, Astrophysics, and Cosmology Center (PITT PACC), University of Pittsburgh, Pittsburgh, PA 15260, USA}
\email{andrewsb@pitt.edu}

\author[0000-0001-8684-2222]{Jeffrey A.~Newman}
\affiliation{Department of Physics and Astronomy, University of Pittsburgh, Pittsburgh, PA 15260, USA}
\affiliation{Pittsburgh Particle Physics, Astrophysics, and Cosmology Center (PITT PACC), University of Pittsburgh, Pittsburgh, PA 15260, USA}
\email{janewman@pitt.edu}

\author[0000-0002-5665-7912]{Biprateep Dey}
\affiliation{Department of Physics and Astronomy, University of Pittsburgh, Pittsburgh, PA 15260, USA}
\affiliation{Pittsburgh Particle Physics, Astrophysics, and Cosmology Center (PITT PACC), University of Pittsburgh, Pittsburgh, PA 15260, USA}
\affiliation{Department of Statistical Sciences, University of Toronto, Toronto, ON M5G 1Z5, Canada}
\affiliation{Canadian Institute for Theoretical Astrophysics (CITA), University of Toronto, Toronto, ON M5S 3H8}
\affiliation{Dunlap Institute for Astronomy \& Astrophysics, University of Toronto, Toronto, ON M5S 3H4, Canada}
\affiliation{Vector Institute, Toronto, ON M5G 0C6, Canada}
\email{biprateep@pitt.edu}

\begin{abstract}
Photometric redshifts (\photozs) are crucial for the cosmology, galaxy evolution, and transient science drivers of next-generation imaging facilities like the Euclid Mission, the Rubin Observatory, and the Nancy Grace Roman Space Telescope.  Previous work has shown that image-based deep learning \photoz\ methods produce smaller scatter than photometry-based classical machine learning (ML) methods on the Sloan Digital Sky Survey (SDSS) Main Galaxy Sample, a testbed \photoz\ dataset.  However, global assessments can obscure local trends.  To explore this possibility, we used a self-organizing map (SOM) to cluster SDSS galaxies based on their $ugriz$ colors.  Deep learning methods achieve lower \photoz\ scatter than classical ML methods for all SOM cells.  The fractional reduction in scatter is roughly constant across most of color space with the exception of the most bulge-dominated and reddest cells where it is smaller in magnitude. Interestingly, classical ML \photozs\ suffer from a significant color-dependent attenuation bias, where \photozs\ for galaxies within a SOM cell are systematically biased towards the cell's mean spectroscopic redshift and away from extreme values, which is not readily apparent when all objects are considered.  In contrast, deep learning \photozs\ suffer from very little color-dependent attenuation bias.  The increased attenuation bias for classical ML \photoz\ methods is the primary reason why they exhibit larger scatter than deep learning methods.  This difference can be explained by the deep learning methods weighting redshift information from the individual pixels of a galaxy image more optimally than integrated photometry.
\end{abstract}

\section{Introduction}
\label{sec:intro}

Next-generation imaging surveys such as the Euclid mission \citep{euclid}, the Vera C.\ Rubin Observatory's Legacy Survey of Space and Time \citep[LSST; ][]{ivezic2019}, and the Nancy Grace Roman Space Telescope \citep[Roman; ][]{spergel2015, akeson2019} will obtain deep multi-band optical and near-infrared imaging over thousands of square degrees.
Nearly all of the extragalactic science drivers of these surveys, such as cosmology, galaxy evolution, and transients, will require redshift estimates \citep{NewmanGruen2022}. Ideally, spectroscopic redshifts (\speczs) would be used to estimate distances, but obtaining spectra for any significant fraction of the photometrically-detected objects is completely infeasible with current facilities \citep[e.g.,][]{Newman2015}. Instead, these imaging surveys must rely on photometric redshifts (\photozs; e.g., \citealt{LsstSRD}).

Photometry-based supervised ML algorithms have produced more precise \photoz\ estimates for imaging surveys than physically-motivated techniques, such as template fitting (which involves fitting galaxy spectral templates to a galaxy's observed integrated photometry), when sufficient training data is available \citep{Salvato2019}. 
More recently, the advent of convolutional neural networks (CNNs) has enabled deep learning \photoz\ models that can leverage pixel-level color information from the images directly (see \citealt{HuertasCompany2023} for a review of deep learning methods applied to galaxy surveys).
These modern image-based deep neural networks produce the current best \photoz\ predictions if given training samples with complete coverage of color--magnitude--redshift space, such as the Sloan Digital Sky Survey (SDSS), even with training sets that are relatively small \citep[e.g., ][]{pasquet2019, hayat2021, Henghes2022, dey2022, treyer2024, AitOuahmed2024}. Despite these successes, it remains unclear what is the dominant mechanism driving their better performance than classical ML methods on global metrics, though previous works have claimed that deep learning algorithms are leveraging morphological information, inclination, and the presence of fainter neighboring galaxies to achieve better results \citep[e.g., ][]{Menou2019, pasquet2019}.

Strong performance on global metrics can mask offsetting systematic errors in local regions of input parameter space.
An extreme case of this effect is the \textsc{trainZ} \photoz\ estimator from the LSST \Photoz\ Data Challenge \citep{Schmidt2020}, which simply predicted the redshift distribution of the training set as the \photoz\ probability distribution function for all objects, regardless of their photometry.
By construction, the \textsc{trainZ} estimator achieved the ideal distribution of the probability integral transform when considered \textit{globally}, beating out the other 11 widely-used \photoz\ methods.
However, it performed the worst in local regions of color space (as characterized by the conditional density estimation loss; see, e.g., \citealt{Izbicki2017}).
A full conditional assessment by regressing across color space (e.g., using the Cal-PIT method of \citealt{calpit}) is beyond the scope of this paper.
Rather, we simply partition color space using a self-organizing map (SOM; \citealt{Kohonen1982, Kohonen2001}; see Section \ref{sec:SOM}) and assess \photoz\ performance in neighborhoods of color space (see \citealt{Jalan2024} for a similar approach, albeit with different goals, conducted contemporaneously but independently).

The paper is organized as follows. Section \ref{sec:data} describes the SDSS dataset used in this analysis, and Section \ref{sec:redshifts} reviews the \photoz\ algorithms used.  Section \ref{sec:SOM} introduces our method of partitioning the galaxy sample in color space with a SOM.  Section \ref{sec:results} presents results on the \photoz\ scatter and bias for individual SOM cells. Finally, Section \ref{sec:dicussion} discusses the contribution of attenuation bias to \photoz\ scatter for classical ML methods.

\section{Data}
\label{sec:data}
Our analysis uses the SDSS \citep{Gunn1998SDSSCamera, Gunn2006SDSSTelescope, YorkEtal2000Sdss, Smee2013SDSSSpectrograph} Main Galaxy Sample (MGS; \citealt{StraussEtal2002SdssMGS}) from Data Release 12 (DR12; \citealt{ AlamEtal2015SdssDR12}).
We started with an identical selection criteria as \citet{pasquet2019}; namely, we selected galaxies based on their dereddened $r$-band Petrosian magnitude ($r \leq 17.8$) and \specz\ ($z_\mathrm{spec} \leq 0.4$), which yielded 516,525 galaxies. 
We split this sample into the same training (464,873 galaxies) and test (51,652 galaxies) sets as \citet{dey2022}.
We further removed galaxies with very faint dereddened model magnitudes ($u > 24$, $g > 20$, $r > 30$, $i > 19$, or $z > 19$) to ensure reliable $ugriz$ photometry.
These cuts reduced our dataset to 509,731 galaxies, with 458,740 galaxies in our training set and 50,991 galaxies in our test set.

We also used galaxy properties from the SDSS photometric catalog \citep{stoughton2002}, such as \texttt{fracDev\_r} (the weight of the de Vaucouleurs component in a combined de Vaucouleurs + exponential profile fit in the $r$-band) and \texttt{expAB\_r} (the semi-minor/semi-major axis ratio for the exponential profile fit in $r$-band, which we note is only reliable for spirals), which we adopt as the bulge-to-total ratio and inclination, respectively.  Throughout this work, we use the bulge-to-total ratio as a single parameter to summarize a galaxy's rest-frame color (leveraging the well-known trend between morphology and rest-frame color for low-$z$ galaxies; e.g., \citealt{blanton2003}), instead of the specific star formation rate, which requires highly accurate spectra that not all surveys can obtain, including SDSS.

\section{Photometric Redshifts}
\label{sec:redshifts}

To compare the differences between photometry-based classical ML \photoz\ methods and image-based deep learning \photoz\ methods, we considered two approaches in each category.  For classical ML methods, we adopted the \photozs\ from the SDSS DR12 catalog as estimated by \citet[][hereafter B16]{beck2016} and our own \photozs\ determined using a random forest.  For the deep learning models, we investigated \photozs\ from \citet{dey2022} and \citet{pasquet2019}.

\subsection{SDSS DR12 Catalog} \label{sec:beck}
For the SDSS DR12 catalog \photozs, \citetalias{beck2016} used local linear regression, which fits a first-order polynomial to a small number of neighbors with labels (i.e., \speczs) in the input space (\texttt{modelMag} $ugriz$ colors and \texttt{cModelMag} $r$ magnitude).
Performing this calculation locally preserves the small-scale structure in the data and allows it to track subtle trends.
\citetalias{beck2016} used 100 neighbors in their fitting procedure, though they implemented more nuanced criteria to handle under-sampled regions of color--magnitude space and to iteratively filter redshift outliers.

\citetalias{beck2016} trained their model using 1,976,978 galaxies with \speczs\ from the entire SDSS DR12 spectroscopic sample---specifically, the MGS, the luminous red galaxy sample \citep{Eisenstein_LGS}, and the Baryon Oscillation Spectroscopic Survey \citep[BOSS;][]{Dawson_BOSS}---plus additional spectroscopy from other surveys, such as DEEP2 \citep{Newman_2013}, VIPERS \citep{Garilli_VIPERS, Guzzo_VIPERS}, and zCOSMOS \citep{Lilly_2009}.  \citetalias{beck2016} estimated \photozs\ for the 208,474,076 galaxies of the SDSS primary photometric catalog, thus their aims were broader than those of the other approaches we consider, which only used MGS galaxies for training and testing.  That being said, MGS galaxies dominate the information content within the color--magnitude--redshift space covered by MGS galaxies, which is further prioritized by the local nature of their regression method.

\subsection{Random Forest}
\label{sec:random_forest}
Inspired by previous random forest \photoz\ efforts \citep[e.g.,][]{carliles2010RFphotoz, CarrascoKind2013, zhou2019}, we trained a random forest \citep{Breiman2001RF} for \photoz\ estimation using the \texttt{RandomForestRegressor} method from the Python package \texttt{Scikit-learn} \citep{PedregosaEtal2011Sklearn} as a classical ML alternative to the local linear regression method of \citetalias{beck2016}. Random forest regression is a simple supervised learning algorithm consisting of a ``forest'' of decision trees that are each trained to predict a value with the final result of the algorithm being the average of each tree's predicted value. To train the algorithm, we used the training set described in Section \ref{sec:data}. We tested providing the random forest with multiple different sets of input features, starting with the features \citetalias{beck2016} used to train their network, as discussed in Section \ref{sec:beck}. To test the impact of input feature choices, we ran random forest models with several combinations of input features:
\begin{enumerate}
    \item \texttt{modelMag} $ugriz$ colors and \texttt{cModelMag} $r$-band magnitude (our fiducial set up);
    \item \texttt{modelMag} $ugriz$ colors, \texttt{cModelMag} $r$-band magnitude, \texttt{modelMag} $ugriz$ color uncertainties, and \texttt{cModelMag} $r$-band magnitude uncertainty;
    \item \texttt{modelMag} $ugriz$ colors, \texttt{modelMag} $r$-band magnitude, the axis ratio for the exponential model (\texttt{expAB\_r}), and the bulge-to-total ratio (\texttt{fracDev\_r}); and
    \item \texttt{modelMag} $ugriz$ colors, \texttt{modelMag} $r$-band magnitude, \texttt{modelMag} $ugriz$ color uncertainties, \texttt{modelMag} $r$-band magnitude uncertainty, the axis ratio for the exponential model (\texttt{expAB\_r}), and the bulge-to-total ratio (\texttt{fracDev\_r}). 
\end{enumerate}
Ultimately, we found only minor differences in \photoz\ performance (both globally and within individual SOM cells) by adding more features, so we chose model 1 (the same features as \citetalias{beck2016}) as our fiducial random forest \photoz\ model.
However, we adopt the \citetalias{beck2016} \photozs\ as our fiducial classical ML \photoz\ method because it produced qualitatively similar results to the random forest approach, but the \citetalias{beck2016} \photozs\ are widely used and are a standard comparison sample for \photoz\ analyses.

\subsection{Deep Learning Models}
\label{sec:encapzulate}

Despite significant differences in architectures, loss functions, and training mechanisms, a variety of modern deep learning models---including CNNs \citep{pasquet2019}, semi-supervised representation learning \citep{hayat2021}, and deep capsule networks \citep{dey2022}---have all significantly out-performed classical ML \photoz\ methods (namely \citetalias{beck2016}) on the SDSS MGS.  In this work, we considered \photozs\ from \citet{pasquet2019} and \citet{dey2022}, both of which were trained using the dataset described in Section \ref{sec:data}. We adopt the latter as our fiducial deep learning \photoz\ model, though we found very similar results with the former.

\subsubsection{Capsule Networks}
\citet{Hinton2011CapsNet} proposed the idea of capsule networks as an alternative to CNNs, where the artificial neurons of a capsule network are organized into groups (capsules), producing the vector analog to the scalar neurons in CNNs such that they are viewpoint equivariant (as opposed to the viewpoint invariance of CNNs) and therefore could be more efficient to train.  
However, actually training capsule networks proved challenging until the introduction of the dynamic routing algorithm by \citet{SabourEtal2017Capsnet} that dramatically improved the viability of their training.  Soon after, \citet{katebi2019} became the first work to apply capsule networks to galaxy images for galaxy morphology prediction.

\citet{dey2022} utilized a deep capsule network for their \encap\ model (compared to the shallow network proposed by \citealt{SabourEtal2017Capsnet}) developed by \citet{RajasegaranEtal2019Deepcaps}.  Its architecture consists of three sub-networks: a deep capsule network for encoding the galaxy images to a low dimensional embedding space, a class-independent decoder network that helps the network generalize with loss terms for image reconstruction and morphological prediction (using disk/spheroid labels from Galaxy Zoo 1; \citealt{Lintott2011GalaxyZoo1}), and a redshift prediction network.  

\subsubsection{Global Photo-z Performance}
\begin{figure*}
{
\includegraphics[width=0.45\textwidth]{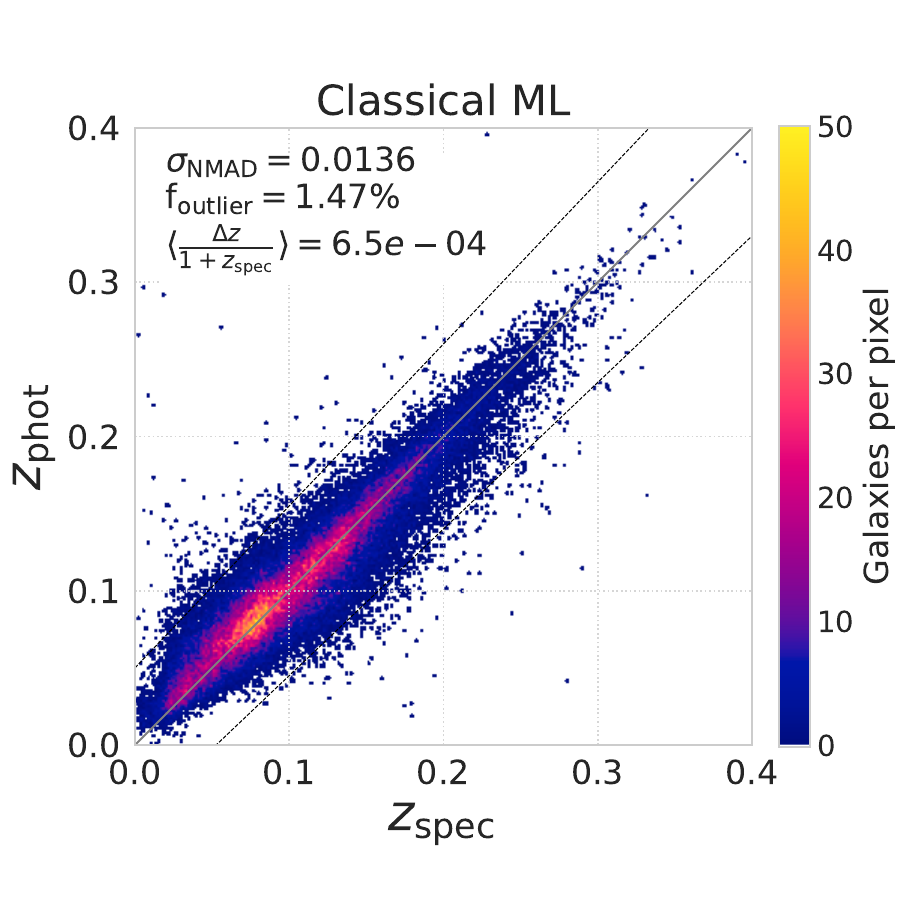}
\hspace{0.5cm}
\includegraphics[width=0.45\textwidth]{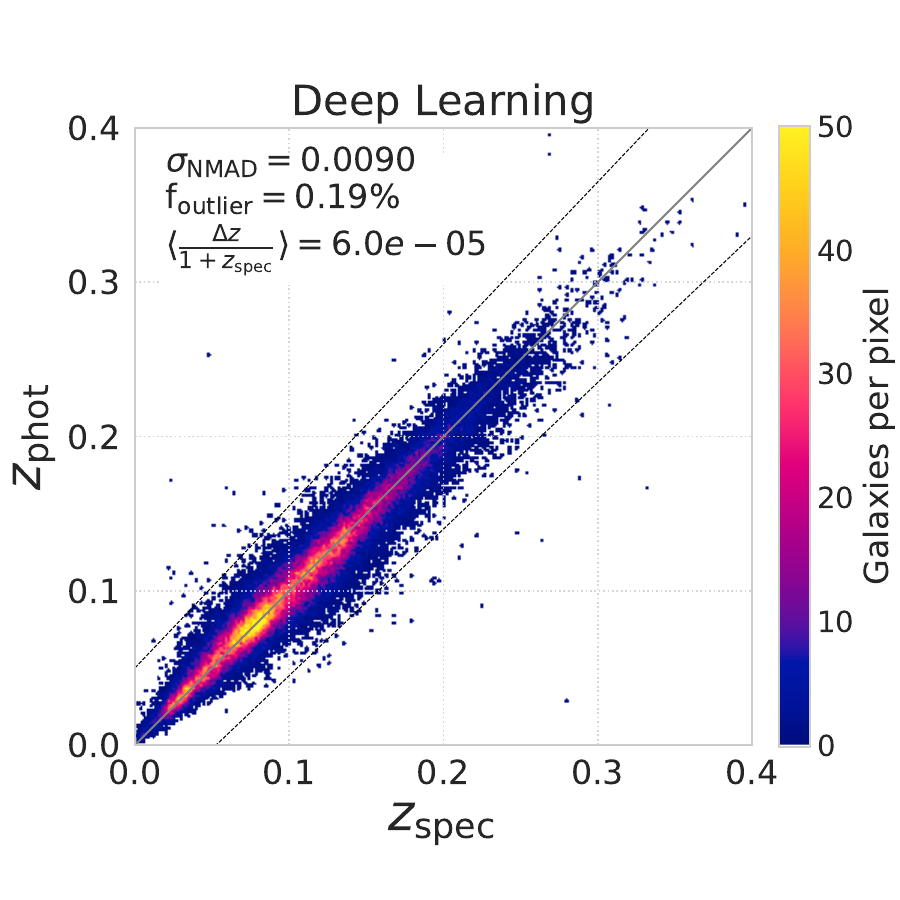}
}
\caption{Global \zphot\ vs.\ \zspec\ for SDSS MGS test set galaxies using the classical ML method of \citetalias{beck2016} (left) and the deep learning model (\encap) of \citet[][right]{dey2022}.  The gray solid line denotes the one-to-one relation.  Points outside of the gray dashed lines ($| \Delta z \, / \,(1+z_\mathrm{spec}) | > 0.05$) are considered outliers.  The \encap\ \photozs\ have smaller global scatter (\nmad), lower fraction of catastrophic outliers (\fout), and a smaller bias ($\langle \Delta z / (1 + z_\mathrm{spec} \rangle$) than the \citetalias{beck2016} \photozs.
We also note that the attenuation bias (see Section \ref{sec:attenuation_bias}) for the classical ML \photozs\ is subtle when the entire dataset is considered, but it manifests as a systematic overestimation of \zphot\ at low \zspec\ and vice versa.
}
\label{fig:global_zphot_zspec}
\end{figure*}

Figure \ref{fig:global_zphot_zspec} shows \zphot\ versus \zspec\ for the \citetalias{beck2016} and \encap\ \photozs\ of SDSS MGS test set galaxies.  Compared to the \citetalias{beck2016} \photozs, the \encap\ \photozs\ have a smaller scatter as measured using the normalized median absolute deviation:
\begin{multline}
    \mathrm{NMAD}(x_\mathrm{true},\, x_\mathrm{predicted}) = 1.4826 \, \times \\
    \mathrm{Median}\left( \left| \frac{\Delta x}{1+x_\mathrm{true}} - \mathrm{Median} \left( \frac{\Delta x}{1+x_\mathrm{true}} \right) \right| \right),
\end{multline}
where $\Delta x = x_\mathrm{predicted} - x_\mathrm{true}$. In the case of \photozs, we define \nmad\ as $\mathrm{NMAD}$(\zspec, \zphot). The \encap\ \photozs\ also have a significantly lower fraction of catastrophic outliers (as defined by $| \Delta z \, / \,(1+z_\mathrm{spec}) | > 0.05$, where $\Delta z = z_\mathrm{phot} - z_\mathrm{spec}$) and a smaller global bias ($\langle \Delta z / (1 + z_\mathrm{spec} \rangle$) than the \citetalias{beck2016} \photozs.  These are generic results of comparing classical ML and deep learning \photoz\ methods for the SDSS dataset, though it is not understood how the deep learning models leverage additional redshift information from the images.

Furthermore, there is a slight but perceptible trend of the \citetalias{beck2016} \photozs\ being biased towards the mean redshift of the dataset, such that they are overestimated at low \zspec\ and underestimated at high \zspec, which is a common bias for ML methods called attenuation bias (see Section \ref{sec:attenuation_bias}).
Below, we compare classical ML and deep learning \photozs\ for sub-samples in small neighborhoods in color space to understand how local trends impact global metrics.

\section{Partitioning Color Space with a Self-Organizing Map}
\label{sec:SOM}
To study the impact of attenuation bias in local neighborhoods of color space, we partition $ugriz$ color space using a SOM.
A SOM is an unsupervised dimensionality reduction algorithm that works by grouping together objects with similar properties into discrete cells, with objects in a cell having properties that are most similar to other objects in the same cell and neighboring cells usually containing galaxies that are broadly similar.  Since SOMs produce easily interpretable results with clear visualizations, they have been used for \photoz\ estimation (e.g., \citealt{CarrascoKindandBrunner2014, Myles2021, Wright2025}) and characterizing the coverage of color space by objects with \speczs\ (e.g., \citealt{Masters2015, Masters2017, Stanford2021, Gruen2023, McCullough2024}). 

For our implementation, we used the \texttt{minisom} python package 
\citep{vettigli2018minisom} to produce a two-dimensional SOM with 10 $\times$ 12 cells using the $u-r$, $g-r$, $r-i$, and $r-z$ colors of all objects in our sample (i.e., including objects in both the training and test sets). Relative to other SOMs, such as the 75 $\times$ 150 cell SOM used in \citet{Masters2015}, our SOM more coarsely bins color space. This choice was motivated by the comparably narrow range in color and redshift of our sample and the need to be able to feasibly perform visual inspection of \photoz\ diagnostic plots for each SOM cell.

\begin{figure*}
\centering
\includegraphics[width=\textwidth]{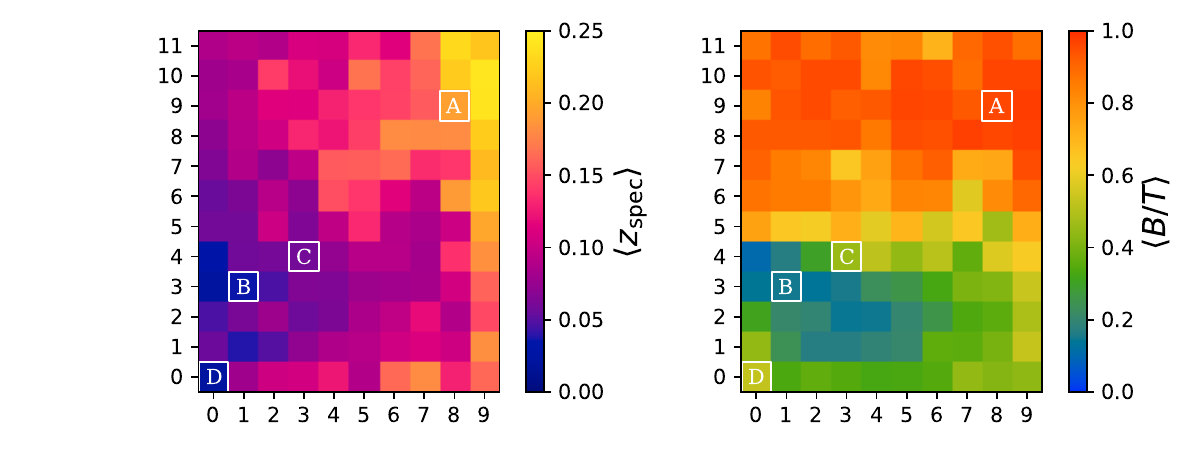}
\caption{A SOM derived from $ugriz$ colors for $\sim$500,000 SDSS MGS galaxies. Cells are color-coded by the mean \specz\ (left panel) and mean bulge-to-total ratio of the objects in each cell (right panel).
Generally, mean \specz\ increases left-to-right, while \meanbt\ increases bottom-to-top.
We highlight four cells: Cells A, B, and C represent typical cells with high, low, and intermediate \meanbt, respectively, 
while Cell D contains very low-$z$ galaxies with a large fraction of catastrophic \photoz\ outliers in the \citetalias{beck2016} catalog.
\label{fig:som}}
\end{figure*}

Figure \ref{fig:som} shows the SOM color-coded by each cell's mean \specz\ (left panel) and mean bulge-to-total ratio (\meanbt; right panel). Generally speaking, the mean \specz\ increases from left to right and \meanbt\ increases from bottom to top.

As representatives of the two main galaxy populations, we chose Cell A (8, 9), which contains red elliptical galaxies (high \meanbt), and Cell B (1, 3), which contains blue disks (low \meanbt). The other two cells of interest show more unusual scenarios. Cell C (3, 4) has an intermediate \meanbt\ and is located in the transition region between low and high \meanbt. It contains a diverse collection of objects, including red disk galaxies \citep{Masters2010}, post-starburst galaxies \citep{Dressler1983}, and galaxies with active galactic nuclei, determined by examining the SDSS and DESI-LS images and SDSS spectra of galaxies in this cell. Finally, Cell D (0, 0) contains very low-$z$ galaxies ($z \lesssim 0.03$), a regime where the \citetalias{beck2016} algorithm suffered from a large fraction of catastrophic outliers \citep{geha2017} but deep learning methods performed well \citep[e.g.,][]{Wu2022, DarraghFord2023}.

\section{Results}
\label{sec:results}

In this section, we compare the scatter and bias of classical ML and deep learning \photozs\ in \textit{localized} regions of color space (i.e., in individual SOM cells) and investigate trends of these quantities with the mean bulge-to-total ratio in each cell (\meanbt).

\subsection{Photo-z Scatter in Individual SOM Cells}
\label{sec:photo_z_sca_per_cell}
\begin{figure}
\centering
\includegraphics[width=\columnwidth]{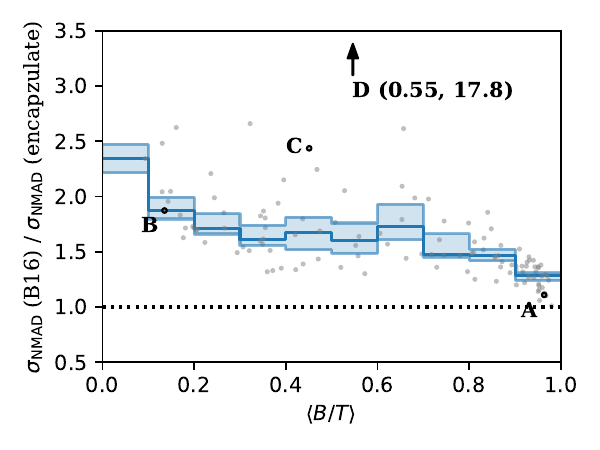}
\caption{The ratio of a robust measure of scatter (\nmad) of \citetalias{beck2016} \photozs\ to the \nmad\ for \encap\ \photozs\ for each SOM cell (gray points) plotted as a function of the cell's mean bulge-to-total ratio (\meanbt).
The solid blue line shows the median \nmad\ ratio computed in bins of \meanbt\ with its uncertainty shown as the shaded region.
The binned median \nmad\ ratio is always greater than $1.0$ (dotted black line). The \nmad\ ratio for Cell D is a major outlier from the general trend due to its extremely high fraction of catastrophic outliers for \citetalias{beck2016} \photozs\ (see Figure~\ref{fig:zphot-zspec-cells}). The scatter for the \citetalias{beck2016} \photozs\ is higher than that of the \encap\ \photozs\ for all cells but approaches the performance of the latter for cells with high \meanbt\ values (e.g., Cell A).
This suggests that deep learning models achieve as good or better \photozs\ compared to classical ML methods across all regions of color space, and they are especially adept at extracting additional redshift information from images of intermediate- and low-\meanbt\ galaxies. 
\label{fig:nmad_ratio_vs_BT}}
\end{figure}

Figure \ref{fig:nmad_ratio_vs_BT} shows the ratio between the \nmad\ values of the \citetalias{beck2016} and \encap\ \photozs\ for individual SOM cells (gray points) as a function of the cell's \meanbt.  For all of the SOM cells, the \nmad\ for the \citetalias{beck2016} \photozs\ is at least as large as that of \encap\ with most of the cells having an \nmad\ value $1.25$--$2.25$ times larger.  The highest \meanbt\ cells typically have smaller \nmad\ from \encap, though a few individual cells (like Cell A) approach parity. The solid blue line and blue shaded regions show the binned median trend and its estimated uncertainty (16$^\mathrm{th}$--84$^\mathrm{th}$ percentile range), respectively, the latter being calculated using a smoothed bootstrap. The first bin (0 $\leq$ \meanbt\ $<$ 0.1) only contained one SOM cell, so we assumed that the uncertainty on its median value would be the average uncertainty of the other bins multiplied by $\sqrt{\pi / 2n}$ (i.e., the standard error on the median), where $n=1$.  The binned medians show a statistically significant trend that is slightly decreasing from \meanbt\ $= 0.1 \rightarrow 0.9$ with an \nmad\ ratio of about $1.5$--$1.7$, which is similar to the global \nmad\ ratio of 1.5.

We found similar results when using the random forest \photozs\ in lieu of the \citetalias{beck2016} \photozs\ and when using the \citet{pasquet2019} \photozs\ instead of the \encap\ \photozs.  This suggests that the deep learning methods are harnessing additional redshift information from the spatially-resolved images that purely photometry-based ML methods cannot access, particularly for SOM cells with low and intermediate \meanbt.

\subsection{Slope of \zphot--\zspec\ in Individual SOM Cells}
\label{sec:conditional_attenuation_bias}

\begin{figure*}
\centering
\includegraphics[width=\textwidth]{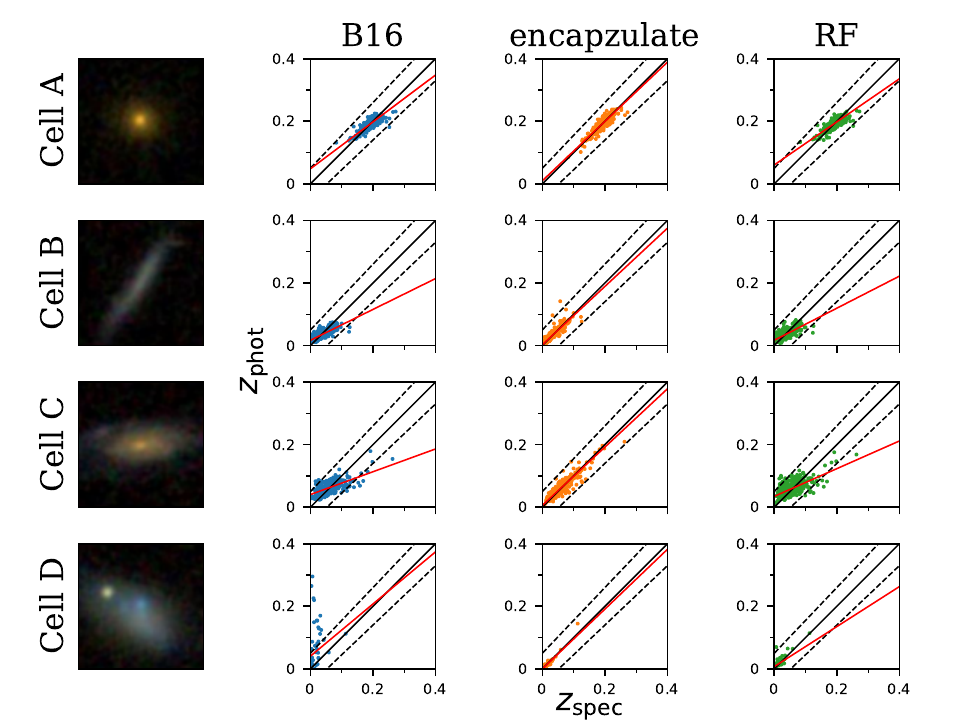}
\caption{For the four SOM cells highlighted in Figure~\ref{fig:som}, we show a 0.42$^\prime$ $\times$ 0.42$^\prime$ DESI-LS $grz$ image of a representative galaxy from the cell (first column), as well as  \zphot--\zspec\ plots for each of three \photoz\ methods: \citetalias{beck2016} (second column), \encap\ (third column), and random forest (fourth column). From top-to-bottom, the rows correspond to Cell A (high \meanbt\ galaxies), Cell B (low \meanbt\ galaxies), Cell C (intermediate \meanbt\ galaxies), and Cell D (very low-$z$ galaxies).
In each \zphot--\zspec\ panel, the solid black line corresponds to the one-to-one relation, the dashed black lines ($\Delta z / (1+z_\mathrm{spec}) = \pm 0.05$) demarcate outliers from typical data, and the red line shows the best-fit line.
Both classical ML learning methods (\citetalias{beck2016} and random forest) yield flatter best-fit lines than the deep learning algorithm (\encap).  The \citetalias{beck2016} \photozs\ for Cell D suffer from a very high catastrophic outlier rate, which is not an issue for the other two methods.
\label{fig:zphot-zspec-cells}}
\end{figure*}

Figure \ref{fig:zphot-zspec-cells} shows $grz$ images (first column) from the Dark Energy Spectroscopic Instrument Legacy Surveys (DESI-LS; \citealt{DESILegacySurveys}) and the \zphot--\zspec\ plots for three \photoz\ methods (\citetalias{beck2016}, \encap, and random forest; second, third, and fourth columns, respectively) with the rows corresponding to Cells A--D (as described in Section \ref{sec:SOM}).
In the \zphot--\zspec\ comparisons, the black solid line denotes the one-to-one line,  and points outside of the black dashed lines are considered outliers using the criterion $| \Delta z \, / \,(1+z_\mathrm{spec}) | > 0.05$.
The red lines show the best-fit lines for each SOM cell determined with least-trimmed-squares regression using the \texttt{LtsFit} package \citep{cappellari2013}. This algorithm works by iteratively clipping outliers in the regression fit (not to be confused with \photoz\ outliers in the \zphot--\zspec\ diagram).
For performing the regression, we adopted the \nmad\ of the objects in a given cell (for a given \photoz\ method) as the estimated \zphot\ uncertainty.
Although \zspec\ uncertainties should be negligible, we assumed an uncertainty of $0.005$ to allow for catastrophic \specz\ failures such as blends or shredded objects.

For Cells A--C, the best-fit lines for the classical ML \photoz\ methods (\citetalias{beck2016} and random forest) are significantly flatter than for the deep learning \photoz\ method (\encap), which instead closely track the one-to-one line in all panels.  In spite of the large fraction of outliers, the best-fit line for the \citetalias{beck2016} \photozs\ in Cell D is close to the one-to-one line because many objects were trimmed in the regression.

\begin{figure}
\centering
\includegraphics[width=\columnwidth]{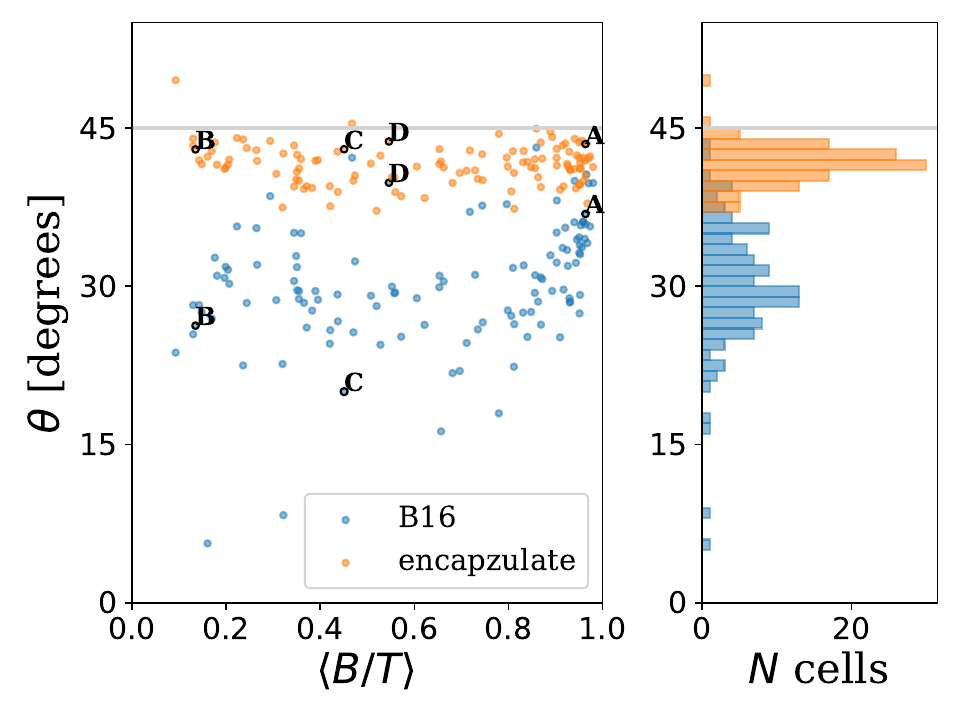}
\caption{Left panel: the angle of the best-fit \zphot--\zspec\ line  from the horizontal axis, $\theta$, as a function of \meanbt\ for each SOM cell. Right panel: a histogram of the number of SOM cells with each $\theta$ value. A perfect correlation would correspond to $\theta$ = 45$^{\circ}$, while a flat best-fit line would have $\theta$ = 0$^{\circ}$.  The \encap\ $\theta$ values are closer to the ideal angle (with a average of 41$^{\circ}$) than \citetalias{beck2016}'s $\theta$ values (average of 29$^{\circ}$) due to the latter suffering from a more severe local attenuation bias.
\label{fig:thetas_vs_BT}}
\end{figure}

We determined the angle of the best-fit \zphot--\zspec\ line by taking the arc-tangent of the slope for each cell, measured counterclockwise from the positive $x$ direction, which we denote by $\theta$. A perfect one-to-one line would correspond to $\theta = 45^{\circ}$.
The left panel of Figure \ref{fig:thetas_vs_BT} shows the resulting $\theta$ values for each SOM cell as a function of \meanbt\ for the \citetalias{beck2016} (blue) and \encap\ (orange) \photozs. The right panel shows the marginal distributions of $\theta$; in both panels, the gray line indicates the ideal ($\theta = 45^{\circ}$) angle.  Due to the attenuation bias, the \citetalias{beck2016} \photozs\ have systematically lower $\theta$ values ($\langle \theta \rangle = 29^{\circ}$) than the \encap\ \photozs\ ($\langle \theta \rangle = 41 ^{\circ}$), which are close to ideal.  We note that the $\theta$ values of the \citetalias{beck2016} \photozs\ approach those of the \encap\ \photozs\ at the very highest \meanbt\ values, mimicking the trend seen in Figure \ref{fig:nmad_ratio_vs_BT}.

The $\theta$ values from our random forest \photozs\ and the \citet{pasquet2019} deep learning \photozs\ are qualitatively similar to those from \citetalias{beck2016} and \encap, respectively, so we have omitted those results from Figure \ref{fig:thetas_vs_BT} for clarity. This suggests that our findings are likely a generic result when comparing image-based deep learning methods to classical ML \photoz\ methods.

\section{Discussion}
\label{sec:dicussion}

\subsection{The Impacts of Attenuation Bias Versus Random Scatter}
\label{sec:attenuation_bias}

\begin{figure*}
\centering
\includegraphics[width=\textwidth]{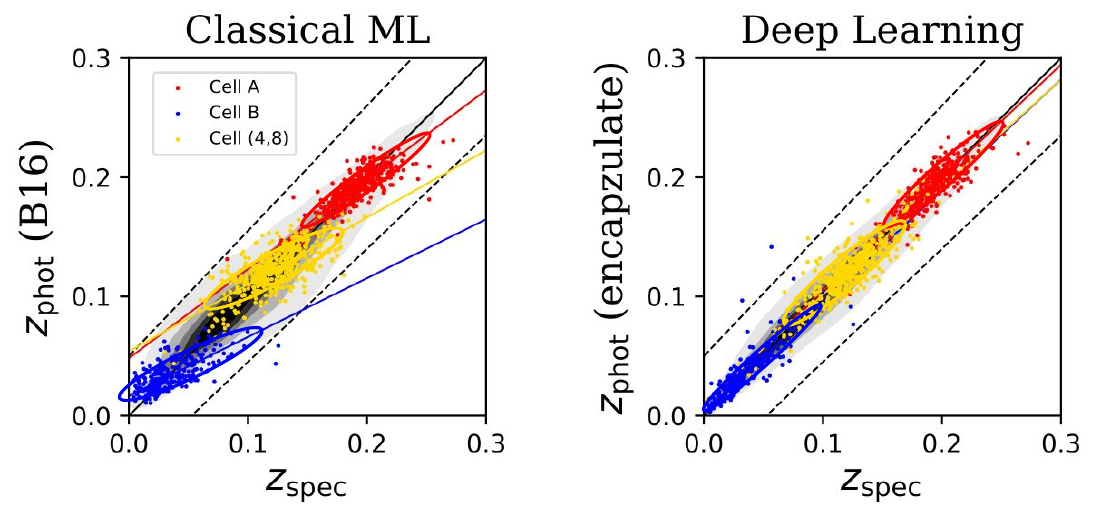}
\caption{Plots of \zphot\ vs.\ \zspec\ for \citetalias{beck2016} (left) and \encap\ (right) \photozs\ showing the full sample as gray contours and objects in selected SOM cells as color-coded points; objects from Cell (4, 8) (gold) are shown in lieu of Cell C for visual clarity.  The best-fit lines for each cell are shown in their respective colors, along with ellipses drawn to highlight the locations of points from each cell.  The significant attenuation bias in individual SOM cells for the \citetalias{beck2016} \photozs\ broadens the global \zphot--\zspec\ distribution; however, this bias is quite small for \encap\ \photozs, resulting in a tighter global \zphot--\zspec\ relation.
\label{fig:zphot_zspec_cellsABC}}
\end{figure*}

As we have shown, photometry-based classical ML \photozs\ exhibit a factor of $\sim$$1.5 \times$ larger scatter than image-based deep learning \photozs\ both when considered globally (see Figure~\ref{fig:global_zphot_zspec}) and in local regions of color space (see Figure~\ref{fig:nmad_ratio_vs_BT}).
Although the global \zphot--\zspec\ trends for both classical ML and deep learning \photozs\ align with the one-to-one line (with  $\theta$ values of  42$^{\circ}$ and 44$^{\circ}$, respectively), classical ML \photozs\ show much flatter slopes in \zphot--\zspec\ plots when restricted to objects in a local neighborhood of color space (e.g., an individual SOM cell) compared to deep learning \photozs, which remain close to the ideal one-to-one line.
For classical ML \photozs, the flatter \zphot--\zspec\ trends combine to form a global \zphot--\zspec\ trend that tracks the one-to-one line but with more scatter than if there were no bias. Figure \ref{fig:zphot_zspec_cellsABC} illustrates schematically how the \zphot--\zspec\ relations for individual color cells combine for classical ML and deep learning \photozs.

The flatter local \zphot--\zspec\ slopes obtained from classic ML methods are a manifestation of a common phenomenon that affects both linear regression and ML methods, known as attenuation bias or regression dilution. This effect causes output values to be biased towards the mean (i.e., with predictions that are too high when values are lowest and too low when values are highest) due to the impact of measurement errors in the independent variable/feature values (e.g., \citealt{fuller1987, akritas1996, loredo2004, kelly2007, Ting2025}).\footnote{While Figure \ref{fig:zphot_zspec_cellsABC} is reminiscent of visualizations of Simpson's paradox \citep{pearson1899, yule1903, simpson1951}, that effect is caused by an unaccounted for confounding variable rather than measurement errors on the independent variable.}
In this section, we assess the relative contributions of attenuation bias and random scatter to the overall scatter within individual SOM cells.

We expect that the overall scatter (within a SOM cell) can be decomposed into two components whose sum in quadrature should be roughly equal to the overall scatter (if standard deviation were used as the metric rather than \nmad, this relationship would be exact).  First, we compute the component of the scatter due to attenuation bias: \sigmaatten\ = $\mathrm{NMAD}$(\zspec, $z_{\mathrm{phot\_best\_fit}}$), where $z_{\mathrm{phot\_best\_fit}}$ is the value of the cell's best fit \zphot--\zspec\ relation at each object's \zspec.  Second, we calculate the scatter of the observed \zphot\ values in a given cell around that cell's best fit \zphot--\zspec\ relation: \sigmascatter\ = $\mathrm{NMAD}$($z_{\mathrm{phot\_best\_fit}}$, \zphot).

\begin{figure*}
\centering
\includegraphics[width=\textwidth]{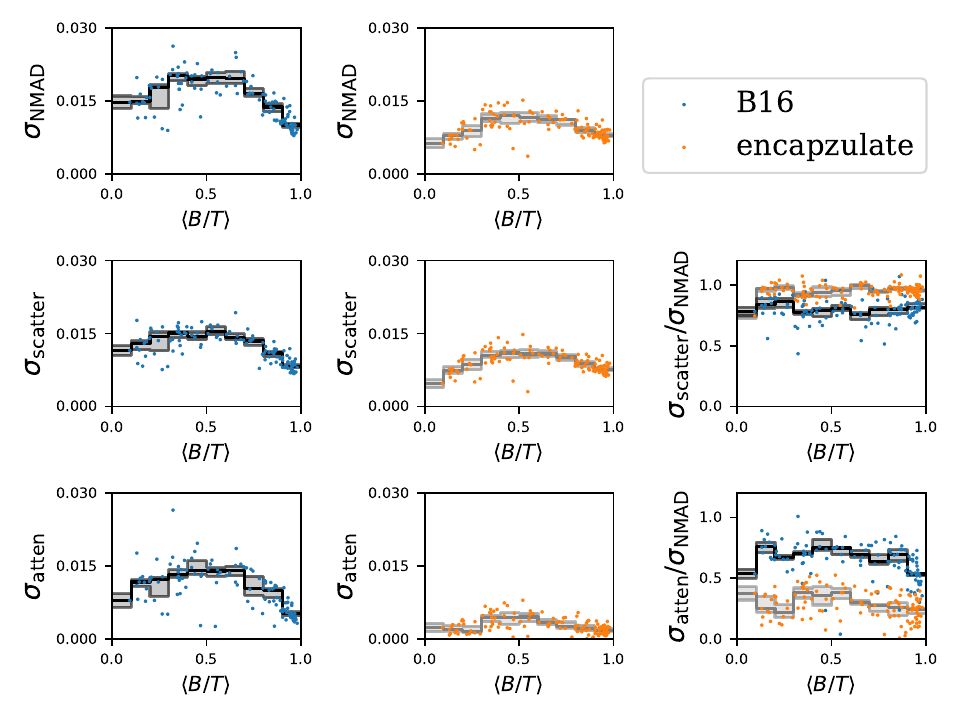}
\caption{
The left (\citetalias{beck2016}; blue points) and middle (\encap; orange points) columns show the overall scatter (\nmad; top row), random scatter (\sigmascatter; middle row), and scatter attributable to attenuation bias (\sigmaatten; bottom row) for individual SOM cells as a function of \meanbt.
The right column shows the ratio of the random scatter (middle row) or attenuation bias (bottom row) to the overall scatter in a cell.
The binned median trends are indicated by the black (\citetalias{beck2016}) and gray (\encap) lines. The shaded regions show the uncertainty in the binned median trends, calculated the same way as in Figure \ref{fig:nmad_ratio_vs_BT}. Most of the reduced overall scatter for \encap\ relative to \citetalias{beck2016} can be attributed to \encap's much lower level of attenuation bias, rather than due to reductions in random errors.
\label{fig:scatters}}
\end{figure*}

Figure \ref{fig:scatters} shows \nmad\ and its two components (\sigmaatten\ and \sigmascatter) for each cell as a function of \meanbt\ for both the \citetalias{beck2016} (left column) and \encap\ (center column) \photozs. The right column of Figure \ref{fig:scatters} shows $\sigma_{\mathrm{scatter}}$/$\sigma_{\mathrm{NMAD}}$ (middle row) and $\sigma_{\mathrm{atten}}$/$\sigma_{\mathrm{NMAD}}$ (bottom row) for each cell and  \photoz\ method. Each panel also shows the binned median trends (black and gray lines) and their uncertainties (dark and light gray regions) with the latter calculated using a smoothed bootstrap as described in Section \ref{sec:photo_z_sca_per_cell}. For the \citetalias{beck2016} \photozs, \sigmaatten\ and \sigmascatter\ provide roughly similar contributions to the overall scatter; however, for the \encap\ \photozs, \sigmaatten\ has a much smaller contribution to the overall scatter than \sigmascatter. In contrast, the \sigmascatter\ values are only a slightly lower for \encap\ than for \citetalias{beck2016}.  We also obtain qualitatively similar results to \encap\ using the \citet{pasquet2019} \photozs, and similar results to \citetalias{beck2016} using our random forest \photozs. Hence, we conclude that the overall lower \photoz\ errors obtained with deep learning methods applied to SDSS can primarily be attributed to their much lower level of attenuation bias, rather than to decreases in random scatter.

\subsection{Deep Learning Leverages Pixels to Mitigate Attenuation Bias}
\label{sec:pixel_noise}

\citet{Ting2025} investigated strategies to mitigate attenuation bias by leveraging many features that are partially correlated because they result from the same underlying physical process (see Sections 2.4.2 and 2.5.2 of that work).
They presented a simple scenario where many noisy features (stellar spectral lines) are linearly correlated with a physical property (stellar elemental abundance).
They found that using many noisy features mitigates attenuation bias relative to predicting from the strongest feature.
However, they did not investigate the most relevant comparison for our use case: specifically, whether attenuation bias differs when regressing from many noisy features (e.g., the set of individual pixels' flux values) versus quantities derived from a combination of these many noisy features (e.g., integrated photometry derived from the pixel-level data).

To develop a more thorough understanding of how the analyses in \citet{Ting2025} relate to our results, we designed a set of Monte Carlo experiments utilizing a toy scenario (see Appendix \ref{sec:appendix} for details).
These experiments allowed us to test the effect of compressing many independent input features (e.g., pixel fluxes) into a single value (e.g., integrated flux), as well as to verify how to apply the predictive framework of \citet{Ting2025} to this use case.
Interestingly, we found that if the relative weightings of features in the combination step are optimal (such that the weighting of a given pixel is determined by how well it can predict an unknown quantity of interest given both its noise level and its degree of physical correlation with that quantity) then the impact of attenuation bias when using individual features for prediction is the same as when using an optimally-weighted combination.
However, if the relative weightings are imperfect, then regressing for the quantity of interest from the individual features yields less attenuation bias---effectively, the regression can approximate the unknown, optimal weighting.

In our case, deep learning \photoz\ models utilize the spectral energy distributions (SEDs) of many individual pixels per object that all share the same redshift (ignoring unassociated blends between foreground and background galaxies, which are very rare at the depths of the SDSS MGS; \citealt{Dawson2016UnrecognizedBlends}) but have uncorrelated noise and can have different SEDs due to different underlying stellar populations\footnote{We note that using multiple measurements of a galaxy's photometry (performed via various methods on the same pixel-level data) would not mitigate attenuation bias because these measurements would be highly redundant and have covariant noise.}. The integrated photometry used for the \citetalias{beck2016} \photozs\ is based upon model-fitting photometry which should (if the model is accurate) weight pixels proportionally to an object's flux.  If the  model used to measure photometry matches the true light distribution of the galaxy perfectly and all pixels' SEDs are equally sensitive to redshift, then it would be an ideal weighting of the pixels and hence would be expected to yield comparable attenuation bias to using all the pixels as independent features for prediction. 

Instead, the attenuation bias in \photozs\ is reduced for predictions using the actual pixel values compared to integrated photometry, particularly for bluer or less bulge-dominated galaxies.  This suggests that the models used to measure photometry for these galaxies are less accurate and/or that some pixels are more informative about redshift than their relative flux implies. Indeed, we would expect that for galaxies with significant spiral structure or irregularity, the simple composite exponential/De Vaucoleurs models used for SDSS photometry would not capture all of the variations in intensity amongst pixels, making the first of these scenarios plausible for later-type galaxies but of little import for early-type galaxies that tend to have much smoother morphologies.  Furthermore, late-type galaxies exhibit much stronger gradients in colors and stellar population properties than early-type galaxies do \citep{deVaucouleurs1961}; the colors of the older, redder populations at the centers of these galaxies are more strongly correlated with redshift than the colors of their outer portions \citep{gladders2000}.  

As a result, both of these explanations for our results are very plausible: later-type galaxies have larger deviations of pixel fluxes from the models used for integrated photometry and also have larger variations in pixel sensitivity to redshift than would be expected from their relative brightnesses in the SDSS photometry models alone.  This can cause redshift predictions from integrated photometry to suffer from greater attenuation bias than predictions from the raw pixel values, especially for galaxies with smaller bulge-to-total ratios, as we have found.  Our results are not sufficient, however, to indicate which of these two possible effects is dominant.  One potential approach for disentangling this puzzle would be to make redshift predictions from photometry integrated within multiple circular or elliptical annuli (which should tend to separate components of different colors; e.g., \citealt{saxena2024}) and compare to the attenuation bias and overall \zphot--\zspec\ relations when predicting redshift from raw pixel values.

A key outstanding question is how the findings from \citet{Ting2025} for attenuation bias in multivariate linear regression should be applied to deep neural networks; we are not aware of any work in the literature directly addressing this issue.
\citet{Ting2025} presented the idea that perhaps attenuation bias for neural networks could be analogous to the linear regression case because neural networks predict an output quantity using a final linear transformation on features extracted through previously applied nonlinear transformations.
However, they showed that the propagation of observed noise through nonlinear transformations (specifically higher-order polynomial regression) is non-trivial, and in fact makes attenuation bias stronger.
This makes sense in the context of the classical bias--variance tradeoff \citep{geman1992} because adding parameters allows a model more flexibility to fit the observations in the training set more accurately (including fluctuations due to the noise), resulting in more severe attenuation bias (as that bias occurs at the ends of the distribution, away from the bulk of the training data).
In this paradigm, the dramatic overparameterization of deep neural networks (i.e., the fact that their number of parameters greatly exceeds the number of data points) would naively suggest that they would suffer from strong attenuation bias.
Instead, we found that deep learning models deliver significantly reduced attenuation bias for \photoz\ estimation.

Considering our results in the context of the ``double descent'' phenomenon \citep{neal2018, belkin2019} suggests a resolution to this tension.
Double descent refers to the fact that across a wide range of deep learning models, there are two separate regimes in which performance improves (i.e., the loss determined from an independent test set decreases) as model capacity (i.e., the number of parameters) increases.
In the classical learning regime (i.e., where the number of model parameters is less than the number of data points), there is an optimal number of parameters for which a model can maximize performance on an unseen test set by balancing the bias--variance tradeoff; the so-called first descent is then the improvement in test set loss that occurs as the number of parameters increases from zero towards this optimal value.
In scenarios where double descent occurs, test set performance is generally poor near the ``interpolation threshold'' where the number of model parameters exactly equals the number of data points, such that models effectively simply memorize the training set \citep{zhang2016DeepLearningGeneralization}.
However, in the modern learning regime where the number of model parameters is far greater than the number of data points, the classical bias--variance tradeoff no longer applies and test set performance once more improves as the number of parameters increases; in this second descent, both bias and variance can decrease simultaneously \citep{belkin2019}.

Like models near the interpolation threshold, deep learning methods have enough parameters to exactly fit the training set.
While the double descent phenomenon is not yet fully understood \citep{sacouto2022DoubleDescentInML}, it is thought that the additional degrees of freedom in deep neural networks allow them to smoothly interpolate between data points in the training set, unlike models near the interpolation threshold, even when trained on noisy data (see, e.g., \citealt{belkin2021}).
From an attenuation bias standpoint, we hypothesize that highly overparameterized neural networks behave more like linear regression than higher-order regression when given many noisy features because of how smoothly they interpolate between training data points, which can significantly mitigate attenuation bias.

\subsection{Outlook}
\label{sec:outlook}

While a full investigation of \textit{how} deep learning \photoz\ methods are able to mitigate the local attenuation bias that affects classical methods is beyond the scope of this work, a better understanding of how deep learning models are able to utilize the  information available in images may provide key insights for designing more efficient \photoz\ strategies, such as new data augmentation procedures for contrastive learning approaches that utilize noise augmentations.  
In particular, these efforts will have important implications for developing deep learning \photoz\ estimation algorithms for Euclid \citep{Siudek2025}, LSST \citep{Merz2025DeepDiscLSST}, and Roman (Khederlarian et al.~in prep.).

\appendix

\section{Monte Carlo Experiments}
\label{sec:appendix}

\citet{Ting2025} found that attenuation bias in linear regression can be mitigated by using many correlated but noisy features as inputs.  However, it was not obvious from their Equation 43 how the predicted severity of attenuation bias would apply to our application of \photoz\ estimation, where we are comparing the utilization of fluxes in individual pixels to integrated photometry (i.e., a weighted/unweighted average of the same pixels).  To develop a greater understanding, we performed a series of Monte Carlo numerical experiments to explore various scenarios about how pixel fluxes are correlated with redshift, which were designed to have  results that would be predictable by Equation 43 of \citet{Ting2025}.

\subsection{Setup}

We generated random data following the framework from Section 2.4.2 of \citet{Ting2025}, which considers the case of multivariate linear regression with correlated features. The dependent variable ($y$ in the \citealt{Ting2025} paper; redshift in our application) is assumed to be determined by the equation $y_\mathrm{true} = \sum_{j=1}^p{\beta_j x_{\mathrm{true},j}}$, where $p$ is the total number of pixels, $\beta_j$ describes the correlation strength between the target quantity and the value of the $j^\mathrm{th}$ pixel, and $x_{\mathrm{true,}j}$ is the true value of that pixel (to which noise is then added, leading to attenuation bias).

To simplify the analysis, \citet{Ting2025} assumed that $x_{\mathrm{true},j} = a_j x_\mathrm{true}$ (i.e., that all the true, noiseless pixel values are linearly dependent upon a single underlying $x_\mathrm{true}$ value for each object/observation) and that $\beta_j = \beta a_j$ (i.e., that the coefficient tying each individual true pixel value to $y_\mathrm{true}$ is proportional to the $a_j$ value for that pixel).  The attenuation bias for a given scenario, $\lambda_p$, then corresponds to the ratio of the recovered value of $\beta$ to the true value.  For simplicity, we set the $x_\mathrm{true}$ values to be evenly distributed between 0.1 and 1.9, with an average value of 1.

We performed our simulations using three different choices for the values of the correlation slopes $a_j$: 
\begin{enumerate}[label=(\alph*)]
\item all $a_j$ values are set identically to 1 (which we will refer to as the ``uniform'' case);

\item the $a_j$ values are uniformly distributed between 0.1--1.9, separated by equal intervals (``deterministic''); or

\item the $a_j$ values are drawn from a uniform random distribution over the interval [0, 2]  (``random'').

\end{enumerate}
In the latter two cases, we divide all coefficients by the mean of the $a_j$ array to ensure that the average $a_j$ value is always one.

In the Monte Carlo simulations, we varied the number of observations/objects used in regression, $n_\mathrm{obs}$ (e.g., 101); the number of features per observation (e.g., 51; i.e., analogous to the number of pixels that contribute to the prediction of a galaxy's redshift in our application); and the Gaussian measurement uncertainty in each feature's value, $\sigma_x$ (e.g., 1).  We added Gaussian-distributed random noise with mean 0 and standard deviation determined by $\sigma_x$ (scaled such that the signal-to-noise ratio for a given pixel is proportional to $a_j$) to all $x_{\mathrm{true},j}$ values to produce a set of observed values for each pixel for all objects in each simulation, which we label $x_{\mathrm{obs},j}$.  Since we are emulating the prediction of redshift, which will have negligible errors when measured spectroscopically, we assume there is no uncertainty in the dependent variable (i.e., $\sigma_y = 0$).

Before performing regression, we calculate both the unweighted mean of the $x_{\mathrm{obs},j}$ values for an observation, $\langle x \rangle$, as well as the optimally weighted mean, $\langle x \rangle_\mathrm{wt} = \frac{\sum_{j=1}^p a_j^2 x_{\mathrm{obs},j}}{\sum_{j=1}^p a_j^2 }$.  We then compare the amount of attenuation bias that results when performing linear regression to predict the value of $y$ from the full set of pixels for each object (which corresponds to the analysis in \citealt{Ting2025}) to the bias produced from regressing for $y$ from $\langle x \rangle$ or $\langle x \rangle_\mathrm{wt}$ values (a scenario which was not considered in \citealt{Ting2025}).

Comparing to our application, regressing from the full set of $x_{\mathrm{obs},j}$ values is analogous to predicting redshift from the full set of pixels in an image.  For the deterministic and random scenarios for $a_j$, predicting from $\langle x \rangle$ is most analogous to predicting $z$ from integrated photometry (i.e., weighting pixels equally, though this approach is suboptimal because pixels with higher signal-to-noise ratios provide more information about redshift).  Predicting from $\langle x \rangle_\mathrm{wt}$ most closely resembles prediction from integrated photometry that weights pixels according to the true light profile of a galaxy (as would be true of model-based photometry that uses a perfect model for each object). However, predicting from $\langle x \rangle_\mathrm{wt}$ will also give greater weight to pixels that are more informative about redshift for reasons other than their flux (e.g., due to having SEDs with stronger spectral breaks), which is not the case for model-based photometry.

In order to correspond directly to the scenarios evaluated in \citet{Ting2025}, all regressions were performed with no constant term.  For the multivariate regression we used the \texttt{OLS} function from the \texttt{statsmodels} Python package; for the cases where we regressed from a single value per object, we instead used the \texttt{linalg.lstsq} routine from \texttt{scipy}.

\subsection{Experiments}
The first test we performed with the Monte Carlo simulations was to investigate how the amount of attenuation bias changed if the $a_j$ values were not all identical.  We tested this by comparing the mean slope across all pixels in the multivariate regression case to the expected value of one; this corresponds to the value of $\lambda_p$ predicted from Equation 43 of \citet{Ting2025}.  We computed this mean for each of 5000 mock datasets and averaged to accurately measure small differences.  We found that our results agreed with the formula to within a fraction of a percent (with the differences likely explained by the finite number of simulations performed): in all cases, the attenuation bias was smaller if the $a_j$ values varied than if they were uniform with the same mean.  This is a consequence of the fact that the quantity $\sum a_j^2$ (which appears in the formula from \citealt{Ting2025}) must be larger than $\sum \langle a_j \rangle^2$ (which is the value of that term for the uniform case, since $\langle a_j \rangle$ is the same in all three of our scenarios for $a_j$) if the variance of the $a_j$ array is nonzero.

The results of this test confirm that the mock datasets conform directly to the scenario presented in Section 2.4.2 of \citet{Ting2025}.  They also illustrate that in situations where different pixels have different amounts of information about redshift (e.g., because of variations in signal-to-noise or SED) we should expect that regressing against the full set of pixel values will yield less attenuation bias than in the case where all pixels are equally informative (with the net amount of information combining all pixels kept constant).

We next tested whether and how the amount of attenuation bias differed if averages were used to make predictions rather than individual pixel values.  For scenario (a) (with equal correlations), we only consider the unweighted average $\langle x \rangle$ (equivalent to the total integrated flux in our application, since they only differ by a factor of the number of pixels averaged); for scenarios (b) and (c), we tested regression of redshift against both the unweighted average and the optimally-weighted average $\langle x \rangle_\mathrm{wt}$ (for scenario (a), the weighted average is identical to the unweighted one).  Since we always normalized the $a_j$ array to have mean one, however, the unweighted averages used in scenarios (b) and (c) should be identical to the values resulting from scenario (a).

For the case where individual pixel values are used, the predicted ratio of the slope with attenuation bias to the true slope, $\lambda_p$, as given by Equation 43 of \citet{Ting2025} reduces to:
\begin{equation}
    \lambda_p = \frac{\sum_{j=1}^{p} a_j^2}{\frac{n_\mathrm{obs} \sigma_x^2}{\sum_{i=1}^{n_\mathrm{obs}} x_{\mathrm{true},i}^2} + \sum_{j=1}^{p} a_j^2}.
\end{equation}
When the averaged values are used in the regression for scenario (a), there is effectively only a single ``pixel'' (i.e., $p=1$), so the expression instead corresponds to:
\begin{equation}
    \lambda_p = \frac{1}{\frac{n_{\rm obs}\sigma_x^2}{\sum_{i=1}^{n_\mathrm{obs}}  x_{\mathrm{true},i}^2} + 1}.
\end{equation}
The results of our Monte Carlo tests agreed with both of these equations closely.  In scenario (a), these equations imply that the attenuation bias will be the same if one regresses against the individual pixel values or if one instead regresses against the simple average of the pixel values; that is indeed what we found.  Furthermore, in scenarios (b) and (c), the attenuation bias that results when an optimally weighted average of the pixel values was used for regression matched the amount of attenuation bias when regressing against the full set of pixel values from each observation.

However, for the case where the correlation strength with the target quantity varied across  pixels, the attenuation bias was less than when the correlations were equal (as was seen in the previous test); for the same reason, the attenuation bias is less when an optimally weighted average is used in scenarios (b) and (c) than when an unweighted average is used.

\subsection{Implications}
These tests have helped to clarify the implications of \citet{Ting2025} for our application.  In particular, they show that we should expect machine learning algorithms that utilize individual pixel values to exhibit less attenuation bias than those that use integrated photometry if both (i) the individual pixels have a varying level of correlation with the quantity we wish to predict (i.e., redshift) and (ii) the integrated photometry does not optimally weight the individual pixel values according to their correlation with that target quantity.

Indeed, in our application we should expect that different pixels have different amounts of intrinsic correlation with redshift, both because of varying signal-to-noise ratios and because of varying stellar populations (which imply that pixel values are not determined by redshift alone).  
In this scenario, for cases where the SED is homogeneous across pixels (as is true in most early-type galaxies), model-based photometry should effectively weight pixels according to their signal-to-noise and hence would be expected to lead to comparable attenuation bias to utilizing all the pixels on their own; however, when the star formation history varies substantially across pixels, that information will not be captured well by the model used for measuring fluxes.  This expectation is consistent with the results of Figure \ref{fig:scatters}, where the reduction of attenuation bias from deep learning methods is much greater for objects with intermediate bulge-to-total ratios (and hence with varying SEDs across pixels) than those that are bulge-dominated and more uniform in color.

As noted by \citet{Ting2025}, when more general machine learning models are used (such as the deep learning models used to predict \photozs) a simple prediction of the amount of attenuation bias is impractical.  However, although the toy model used in this section is obviously unrealistic, the insights gained do suggest a very plausible explanation for our results.

\begin{acknowledgments}
E.R.M., B.H.A., J.A.N., and B.D.\ acknowledge the support of the National Science Foundation under Grant No.~AST-2009251. Any opinions, findings, and conclusions or recommendations expressed in this material are those of the author(s) and do not necessarily reflect the views of the National Science Foundation.  This work was supported by a NASA Pennsylvania Space Grant Consortium (PSGC) grant to the University of Pittsburgh and by cost-share funding from the Dietrich School of Arts and Sciences at the University of Pittsburgh (Award Number S000978-NASA, Prime Award Number 80NSSC20M0097).  B.D.\ is a postdoctoral fellow at the University of Toronto in the Eric and Wendy Schmidt AI in Science Postdoctoral Fellowship Program, a program of Schmidt Sciences.  We thank Yuan-Sen Ting, Maciej Bilicki, Andrew Engel, and Tom Loredo for insightful conversations.  We also appreciate the anonymous referee, whose thoughtful comments materially improved this work.

This research used resources of the National Energy Research Scientific Computing Center, a DOE Office of Science User Facility supported by the Office of Science of the U.S.\ Department of Energy under Contract No.\ DE-AC02-05CH11231 using NERSC award HEP-ERCAP0022859 and HEP-ERCAP0033572. 

Funding for SDSS-III has been provided by the Alfred P.~Sloan Foundation, the Participating Institutions, the National Science Foundation, and the U.S.\ Department of Energy Office of Science. The SDSS-III website is http://www.sdss3.org/.

SDSS-III is managed by the Astrophysical Research Consortium for the Participating Institutions of the SDSS-III Collaboration including the University of Arizona, the Brazilian Participation Group, Brookhaven National Laboratory, Carnegie Mellon University, University of Florida, the French Participation Group, the German Participation Group, Harvard University, the Instituto de Astrofisica de Canarias, the Michigan State/Notre Dame/JINA Participation Group, Johns Hopkins University, Lawrence Berkeley National Laboratory, Max Planck Institute for Astrophysics, Max Planck Institute for Extraterrestrial Physics, New Mexico State University, New York University, Ohio State University, Pennsylvania State University, University of Portsmouth, Princeton University, the Spanish Participation Group, University of Tokyo, University of Utah, Vanderbilt University, University of Virginia, University of Washington, and Yale University.
\end{acknowledgments}

\software{
Colorcet \citep{Kovesi2015Colorcet},
LtsFit \citep{cappellari2013},
Matplotlib \citep{Hunter2007Matplotlib},
MiniSom \citep{vettigli2018minisom},
Numpy \citep{harris2020Numpy},
Pandas \citep{mckinney2010Pandas, reback2020Pandas},
Scikit-Learn \citep{PedregosaEtal2011Sklearn},
Scipy \citep{VirtanenEtal2020Scipy}, and
Statsmodels \citep{statsmodels}.
}

\bibliography{library}{}
\bibliographystyle{aasjournal}

\end{document}